\documentclass{jfm}
\usepackage{graphicx}
\usepackage[]{psfrag}
\usepackage{xcolor}
\usepackage{subfig}
\usepackage{tikz}
\usepackage{interval}

\newcommand{\SF}[1][]{
  \ensuremath{S^{\theta}_{#1}(r) } }
  
  \newcommand*{\shifttext}[2]{%
  \settowidth{\@tempdima}{#2}%
  \makebox[\@tempdima]{\hspace*{#1}#2}%
}
  
\shorttitle{ESS scaling of temperature structure functions in RB turbulence}
\shortauthor{D. Krug and others}

\title{Transition to ultimate  Rayleigh-B\'{e}nard turbulence revealed through extended self similarity scaling analysis of the temperature structure functions}

\author{Dominik Krug\aff{1,2}
  \corresp{\email{d.j.krug@utwente.nl}},
  Xiaojue Zhu\aff{1}, 
  Daniel Chung\aff{2},
  Ivan Marusic\aff{2},
  Roberto Verzicco \aff{1,3},
 \and Detlef Lohse\aff{1,4}}

\affiliation{
\aff{1}Physics of Fluids Group and Twente Max Planck Center, 
Department of Science and Technology, Mesa+ Institute,
 and J.M. Burgers Center for Fluid Dynamics, University of Twente, P.O Box 217, 7500 AE Enschede,  The Netherlands \\
\aff{2}Department of Mechanical Engineering, University of Melbourne, 3010,Victoria, Australia\\
\aff{3}Dipartimento di Ingegneria Industriale, University of Rome `Tor Vergata', Via del
Politecnico 1 Roma 00133, Italy\\
\aff{4}Max Planck Institute for Dynamics and Self-Organization, 37077 G\"ottingen, Germany}

\begin{document}

\maketitle

\begin{abstract}
In turbulent Rayleigh-B\'{e}nard (RB) convection, a transition to the so-called ultimate 
regime, in which the boundary layers (BL) are of turbulent type, has been postulated. Indeed, at very large Rayleigh  number  $Ra \approx 10^{13}-10^{14}$ a transition in the scaling of the global Nusselt number $Nu$ (the dimensionless heat transfer) and the Reynolds number  with  $Ra$ has been observed in experiments and very recently in direct numerical simulations (DNS) of two-dimensional (2D) RB. In this paper we analyse the local scaling properties of the lateral temperature structure functions in the BLs of this simulation of 2D RB, employing extended self-similarity (ESS) (i.e., plotting the structure functions against each other, rather than only against the scale) in the spirit of the attached eddy hypothesis, as we have recently introduced for velocity structure functions in wall turbulence (Krug \textit{et al.}, \textit{J. Fluid Mech.}, vol. 830, 2017, pp. 797-819). We find no ESS scaling below the transition and in the near wall region. However, beyond the transition and for large enough wall distance $z^+ >  100$, we find clear ESS behaviour, as expected for a scalar in a turbulent boundary layer.  In striking correspondence to the $Nu$ scaling, the ESS scaling region is negligible at $Ra = 10^{11}$ and well developed at  $Ra = 10^{14}$, thus providing strong evidence that the observed transition in the global Nusselt number at $Ra \approx 10^{13}$ indeed is the transition from a laminar type BL to a turbulent type BL. Our results further show that the relative slopes for scalar structure functions in the ESS scaling regime are the same as for their velocity counterparts extending their previously established universality. The findings are confirmed by comparing to scalar structure functions in 3D turbulent channel flow.

\end{abstract}


\section{Introduction}
\begin{figure} 
\psfrag{a}[c][c][1]{$(a)$}
\psfrag{b}[c][c][1]{$(b)$}
\psfrag{p}[c][c][1]{$p$}
\psfrag{Nu}[c][c][1]{$Nu/Ra^{0.35}$}
\psfrag{Ra}[c][c][1]{$Ra$}
\centering
\vspace{0.5cm}
\begin{tikzpicture}
\node[anchor=south west,inner sep=0] (image) at (-5.,0) {{\includegraphics[scale = 1]{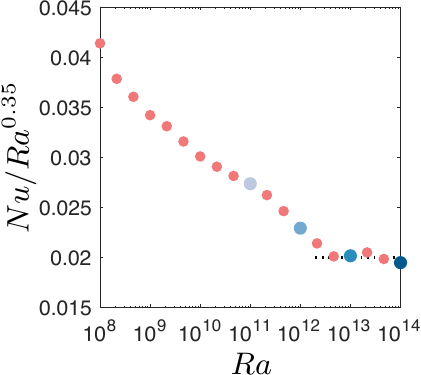}}};
\begin{scope}[x={(image.south east)},y={(image.north west)}]
  \node[inner sep=0,anchor=center] (note2) at (-5.6,1.02) {\scalebox{1}{$(a)$}};
 \end{scope}

        \node[anchor=south west,inner sep=0] (image) at (-0.4,2.25) {{\includegraphics[width = 0.67\textwidth,
        trim= 1.2cm 0.85cm 1.3cm 0.8cm, clip]{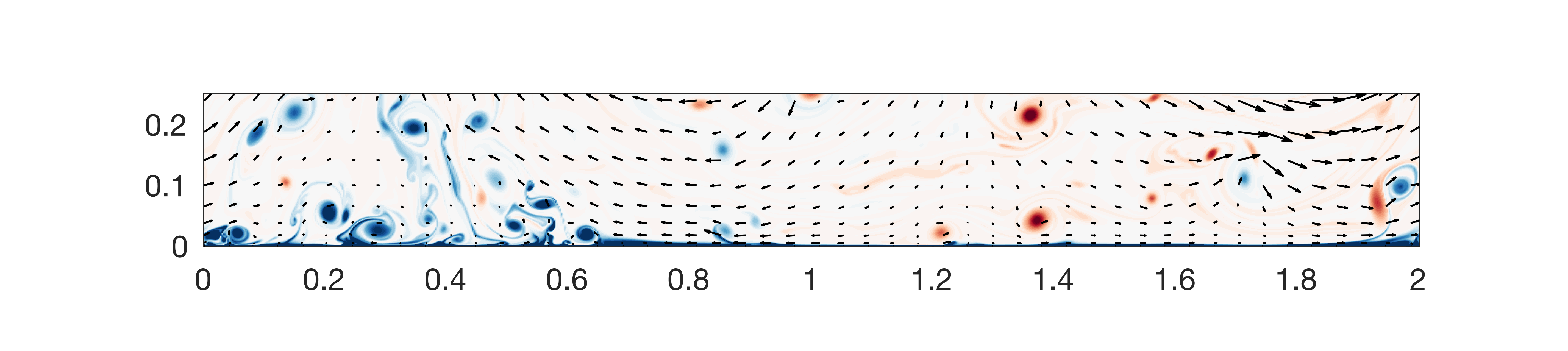}}};
        \begin{scope}[x={(image.south east)},y={(image.north west)}]
  \node[inner sep=0,anchor=center] (note2) at (-0.02,0.85) {\scalebox{1}{$z$}};
                 \node[inner sep=0,anchor=center] (note1) at (0.14,0.95) {\scalebox{0.9}{$Ra = 10^{11}$}};       
                                  \node[inner sep=0,anchor=center] (note1) at (0.98,0.395) {\scalebox{0.9}{$(b)$}};        \end{scope}
         \node[anchor=south west,inner sep=0] (image) at (-0.4,0.3) {\includegraphics[width = 0.67\textwidth,
         trim= 1.2cm 0.45cm 1.3cm 0.8cm, clip]{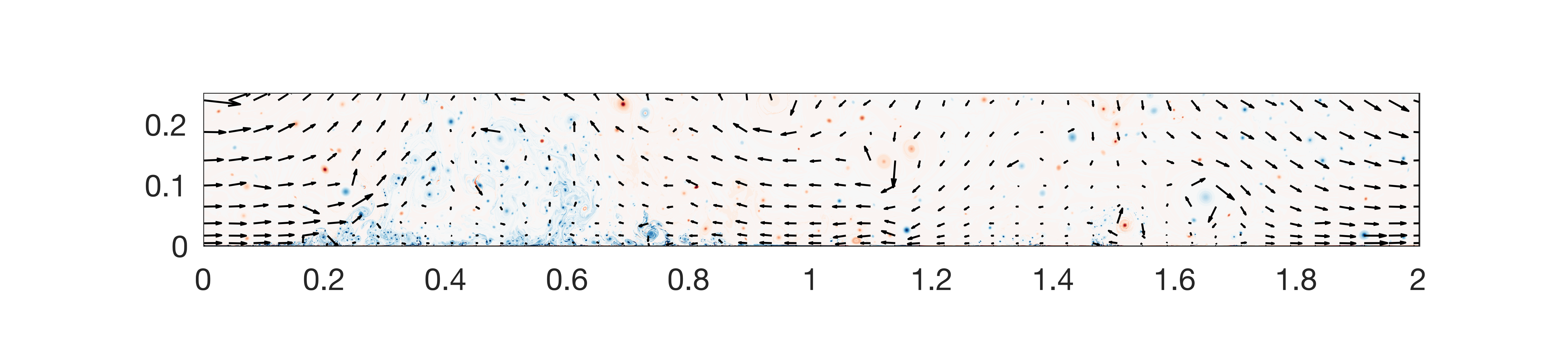}};
        \begin{scope}[x={(image.south east)},y={(image.north west)}]
                \node[inner sep=0,anchor=center] (note1) at (0.478,-0.05) {\scalebox{1}{$x$}};
                 \node[inner sep=0,anchor=center] (note2) at (-0.02,0.7) {\scalebox{1}{$z$}};
                 \node[inner sep=0,anchor=center] (note1) at (0.15,1.05) {\scalebox{0.9}{$Ra = 10^{14}$}};
                 \node[inner sep=0,anchor=center] (note1) at (1,0.9) {\scalebox{0.9}{$(c)$}};
        \end{scope}
    \end{tikzpicture}

\caption{\label{fig:Nu}a) Reduced Nusselt number $Nu/Ra^{0.35}$ from \citet{Zhu2018}; blue symbols indicate the $Ra$ values investigated here. Instantaneous snapshots of temperature (contours for  $0.3 \leq \theta/\Delta \Theta \leq 0.7$ from blue to red) with rescaled velocity vectors at $Ra = 10^{11}$ (b) and $Ra = 10^{14}$ (c). Note that only a fourth of the vertical domain ($z$-direction) is shown.}
\end{figure}

Thermal convection is relevant to a wide range of applications across various fields such as building ventilation \citep[e.g.][]{Linden1999}, atmospheric \citep[e.g.][]{Hartmann2001} or oceanic \citep[e.g.][]{Rahmstorf2000} flows.  The  phenomenon is widely studied in terms of the paradigmatic case of Rayleigh-B\'{e}nard (RB) convection \citep[see the reviews of ][]{Ahlers2009, Lohse2010,Chilla2012}, in which two horizontal plates at a distance $L$ are cooled from above and heated from below. In the presence of a gravitational acceleration $g$, a flow is driven with properties (for given Prandtl number $Pr \equiv \nu/\kappa$ and aspect ratio $\Gamma = D/L$)  that are controlled by the Rayleigh number $Ra \equiv \alpha g {\Delta} \Theta/(\nu \kappa)$. Here, $\nu$ and $\kappa$ are kinematic viscosity and the thermal expansion coefficient, respectively, $D$ is some lateral length scale and $\Delta \Theta \equiv \Theta_b -\Theta_t$ denotes the temperature difference between the bottom (held at $\Theta_b$) and top ($\Theta_t$) plates. The primary response parameter of the system is the resulting heat flux, which in its non-dimensional form is given by the Nusselt number $Nu$ relating the actual heat transfer to the one in the purely conductive case. 
The behaviour of the system at very high $Ra$ is of interest in many applications, and theoretical work \citep{Kraichnan1962,Spiegel1971,Grossmann2000,Grossmann2011} predicts the existence of a so called  `ultimate regime', in which the scaling $Nu \propto Ra^{\beta}$ switches from the classical $\beta \leq 1/3$ \citep{Malkus1954} to  $\beta > 1/3$. This transition is related to a change of the boundary layer (BL) structure, of both velocity and temperature, from a laminar to turbulent type. A transition in the $Nu$ scaling  ---as well as in the Reynolds number scaling--- has indeed been observed experimentally, most convincingly by \citet{He2012transition,He2012heat,He2015} and very recently also in a direct numerical simulation (DNS) of two-dimensional (2D) RB by \citet{Zhu2018}.   Results of the latter are reproduced in figure \ref{fig:Nu}a, where the transition is evident from the change in slope of the (compensated) $Nu$ at $Ra \approx 10^{13}$. As expected from the theory and indicative of the turbulent nature of the BLs, \citet{Zhu2018} found logarithmic dependencies with respect to the distance from the wall for the mean temperature and velocity BLs. In the following, we will characterize the thermal BLs further by studying their lateral structure functions.

 Such an analysis originated from \citet{Davidson2006}, who pointed out that the $k^{-1}$ spectral scaling ($k$ being the streamwise wavenumber) of the streamwise velocity power spectrum predicted by the attached-eddy hypothesis \citep{Townsend1976,Perry1982}, is equivalent to ---and more readily observed as--- a $\ln(r/z)$ scaling of the second order longitudinal structure function. Here and in the following, $r$ denotes the streamwise separation distance and $z$ is the distance off the wall. \citet{deSilva2015scaling} found that the $\ln(r/z)$ scaling also applies to velocity structure functions of arbitrary even order $2p$ such that for the so-called energy-containing range $z<r\ll \delta$ ($\delta$ being the boundary layer thickness)
 \begin{equation}
 S_p^u\equiv  \langle \Delta u ^{2p}\rangle^{1/p}/U_{\tau}^2 = E_p^u +D_p ^u\ln{\frac{r}{z}}.
 \label{eq:Sdir}
 \end{equation}
Here $\Delta u (r,z)$ is the velocity increment between two points at a distance $z$ off the wall separated by $r$ along the streamwise direction, $U_{\tau}$ is the mean friction velocity and $E_p^u$, $D_p^u$ are constants. We use superscript $u$ to denote quantities relating to the velocity and $\theta$ when referring to the scalar later on. While the direct scaling according to (\ref{eq:Sdir}) is only observed at relatively large Reynolds numbers $Re_{\tau} =\delta U_{\tau}/\nu \sim O(10^4)$,  \cite{deSilva2017} and \cite{Krug2017ESS} have shown that a relative scaling is evident at much smaller $Re_{\tau}$ if the so called extended self-similarity (ESS) framework is employed. In this case, the scaling is not analysed as a function of $r$ but --- in the spirit of the original ESS hypothesis by \citet{Benzi1993,Benzi1995} --- relative to a structure function of different order. For an arbitrary reference order $2m$ this results in the 'ESS form' of  (\ref{eq:Sdir}), namely
 \begin{equation}
 S_p^u=\frac{D_p^u}{D_m^u}S_m^u+E_p^u-\frac{D_p^u}{D_m^u}E_m^u.
 \label{eq:ESS}
 \end{equation}
 \citet{deSilva2017} and \citet{Krug2017ESS} found that the linear scaling of (\ref{eq:ESS}) could not only be observed at relatively low $Re_{\tau}$ well within the capabilities  of current DNS, but also that the relative slopes $D_p^{u}/D_m^{u}$ exhibit non-trivial, i.e. non-Gaussian, universal behaviour across various flow geometries such as flat plate boundary layers, pipe and channel flow and even Taylor-Couette flow. From our experience in Taylor-Couette flow \citep{Krug2017ESS} we learned that ESS-scaling according to (\ref{eq:ESS}) is not observed if large scale structures in the bulk (such as the Taylor rolls) contribute to the velocity component under investigation. This is also the case for the wall parallel velocity component in RB, as can be seen from the snapshots in figures \ref{fig:Nu}(a,b). Since no equivalent scaling exists for the wall-normal component \citep{Perry1982}, the 2D velocity field in RB is ruled out as a suitable candidate for this scaling analysis. Therefore the current study instead focuses on structure functions of the temperature field
\begin{equation}
\SF[p]\equiv   \langle \Delta \theta^{2p}\rangle^{1/p}/\Theta_\tau^2 .
\end{equation}
Here, $\Delta \theta$ and $\Theta_{\tau}=-\kappa\partial_z\Theta|_{z=0}/U_{\tau}$ are analogous to $\Delta u$ and $U_{\tau}$, respectively.  ESS-scaling of scalar structure functions in the energy containing range has not been considered yet. It is however conceivable that the situation will be similar to velocity structure functions: 
The theory underlying (\ref{eq:Sdir}) and (\ref{eq:ESS}) is based on an inertial assumption, which implies that momentum transport is dominated by turbulent eddies that are larger than and therefore constrained by $z \gg \eta$ (where $\eta$ is the Kolmogorov length scale) but smaller than $\delta$. In the spirit of the Reynolds analogy and for $Pr \approx 1$ these eddies can be assumed to affect momentum and scalar similarly.  Consequently, the relevant arguments based on the attached eddy hypothesis \citep[see][]{deSilva2015scaling,Krug2017ESS}  can be expected to transfer to the scalar field as well. 
It is therefore our objective here to investigate whether and at what $Ra$ the thermal boundary layers in RB convection exhibit an ESS scaling according to 
 \begin{equation}
 S_p^{\theta}=\frac{D_p^{\theta}}{D_m^{\theta}}S_m^{\theta}+E_p^{\theta}-\frac{D_p^{\theta}}{D_m^{\theta}}E_m^{\theta}.
 \label{eq:ESSscalar}
 \end{equation}
The expectation is that such a scaling regime should originate coinciding with the $Nu$-scaling transition, which has been link to the emergence of logarithmic boundary layers \citep{Grossmann2011}. Before moving ahead, we would like to mention that the attached eddy framework, from which (\ref{eq:Sdir}) and (\ref{eq:ESS}) can be derived, 
 is not new in the RB context. It has already been referred to by \citet{Ahlers2014}, who investigate logarithmic dependencies of temperature profiles, and  by \citet{He2014} in a study of $f^{-1}$ temperature power spectra scalings the spatial equivalent to the $k^{-1}$ scaling) in order to interpret the observations made.

Since the data set of \citet{Zhu2018} that will be employed for this endeavour  is 2D, we will also check our results by using a three dimensional channel simulation as a reference and for comparison.  A brief overview over both datasets will given in \S\ref{sec:data}, before presenting our results in \S \ref{sec:results} and conclusions in \S\ref{sec:conc}.
 
\section{Datasets}\label{sec:data}
The simulations of \citet{Zhu2018} where performed using the second-order finite-difference code AFiD \citep{vdPoel2015} on a two-dimensional domain with periodic boundary conditions in the lateral direction with $\Gamma =2$ and $Pr = 1$. While the original dataset spans six decades $Ra\in\interval[]{10^8}{10^{14}}$, only four data points for $Ra \geq 10^{11}$ are used here (see figure \ref{fig:Nu}a). We refer to the original publication for further details on the numerical setup and validation of the results. One of the particularities of the 2D setup is that the large scale structures, which can be observed in the vector maps of figures \ref{fig:Nu}(a,b), remain almost fixed in place allowing for simple temporal averaging. Since the temporal mean velocity gradient changes sign along the plate, we take the spatial mean of its absolute value when computing $U_{\tau}$ in RB convection. 
For the present simulations we obtain $L^+/2 = LU_\tau/(2\nu) \approx \left[2400; 5700; 12300; 34400\right]  $ at $Ra =[10^{11};10^{12};10^{13};10^{14}]$, respectively.

A DNS of channel flow was performed using the fourth-order code described in \citet{Chung2014} at  $Re_{\tau}=hU_\tau/\nu=590$, where $h$ is half the channel height. The periodic (in streamwise and spanwise directions) box  of size $12h \times 4h \times 2h$ was discretised  by $640\times 320\times 240$ grid points in the streamwise, spanwise, and wall-normal directions, respectively. A  passive scalar with $Pr=1$ was  added with values fixed to $-0.5$ at the bottom and $0.5$ at the top wall as boundary conditions.

 Convergence of the statistics computed from both datasets was checked by plotting the premultiplied probability density function of $\Delta \theta$ at various $r$ and found acceptable up to tenth order.
 
\section{Results}\label{sec:results}
\subsection{Direct analysis of the scaling in the energy-containing range in RB convection}\label{sec:sfun}
\begin{figure} 
\psfrag{a}[c][c][1]{$(a)$}
\psfrag{b}[c][c][1]{$(b)$}
\psfrag{c}[c][c][1]{$(c)$}
\psfrag{d}[c][c][1]{$(d)$}
\psfrag{e}[c][c][1]{$(e)$}
\psfrag{f}[c][c][1]{$(f)$}
\psfrag{g}[c][c][1]{$(g)$}
\psfrag{h}[c][c][1]{$(h)$}
\psfrag{i}[c][c][1]{$(i)$}
\psfrag{data1}[l][l][0.8]{$Ra = 10^{11}$}
\psfrag{data2}[l][l][0.8]{$Ra = 10^{12}$}
\psfrag{data3}[l][l][0.8]{$Ra = 10^{13}$}
\psfrag{data4}[l][l][0.8]{$Ra = 10^{14}$}
\psfrag{S1}[c][c][1]{$S_1^\theta$}
\psfrag{S2}[c][c][1]{$S_2^\theta$}
\psfrag{S5}[c][c][1]{$S_5^\theta$}
\psfrag{ronz}[c][c][1]{$r/z$}
\psfrag{DS1}[c][c][0.9]{$\Delta S_1^\theta$}
\centering
\vspace{0.5cm}
\subfloat{\begin{tikzpicture}
        \node[anchor=south west,inner sep=0] (image) at (0,0) {\hspace{0.45cm}\includegraphics[scale =1]{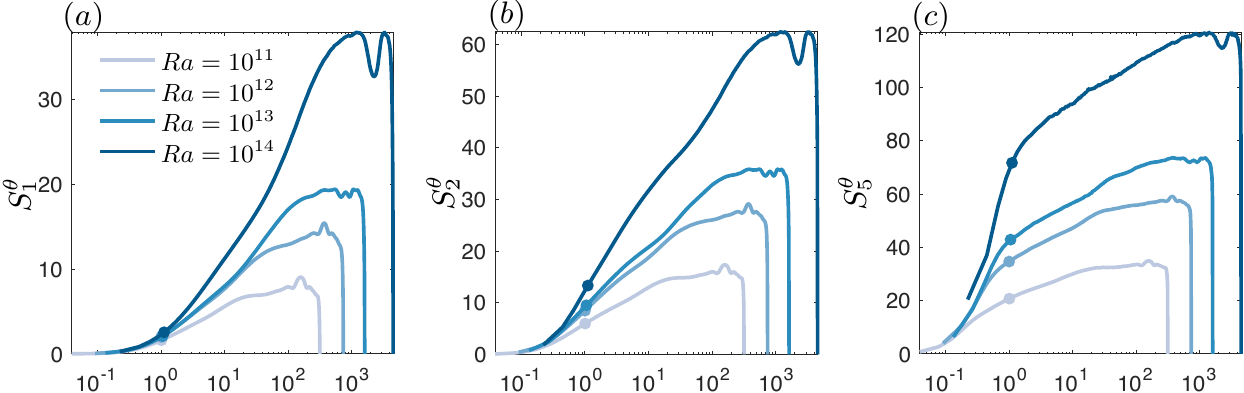}};
        \begin{scope}[x={(image.south east)},y={(image.north west)}]
                \node[inner sep=0,anchor=center] (note1) at (0,1.01) {\scalebox{0.9}{$z^+=30$}};
        \end{scope}
    \end{tikzpicture}}\\
    \subfloat{\begin{tikzpicture}
        \node[anchor=south west,inner sep=0] (image) at (0,0) {\hspace{0.4cm}\includegraphics[scale = 1]{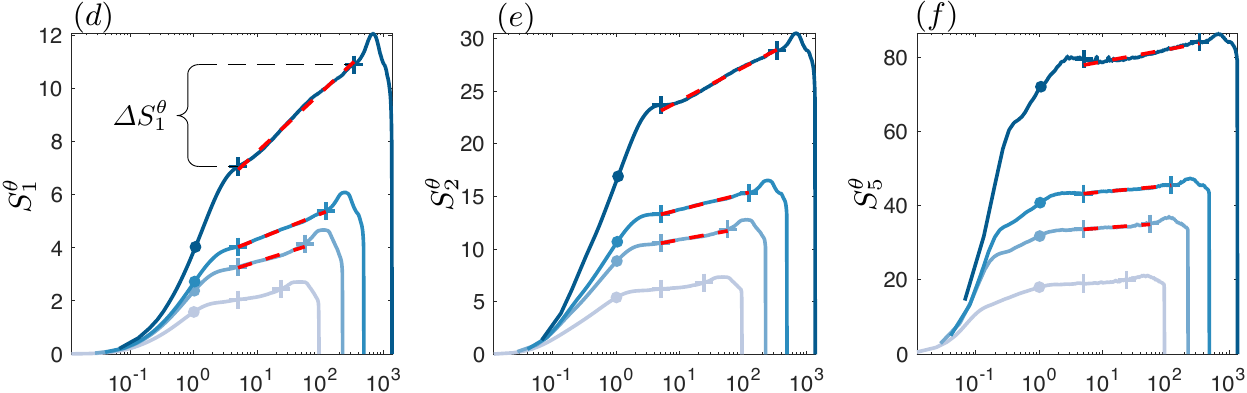}};
        \begin{scope}[x={(image.south east)},y={(image.north west)}]
                \node[inner sep=0,anchor=center] (note1) at (0,1.01) {\scalebox{0.9}{$z^+=100$}};
        \end{scope}
    \end{tikzpicture}}\\
    \subfloat{\begin{tikzpicture}
        \node[anchor=south west,inner sep=0] (image) at (0,0) {\hspace{0.35cm}
        \includegraphics[scale = 1]{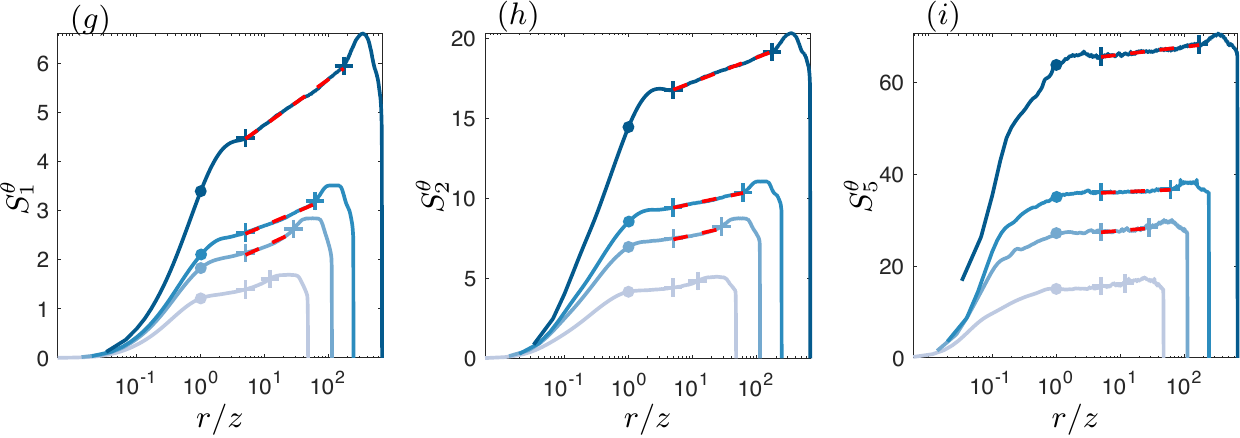}};
        \begin{scope}[x={(image.south east)},y={(image.north west)}]
                \node[inner sep=0,anchor=center] (note1) at (0,1.01) {\scalebox{0.9}{$z^+=200$}};
        \end{scope}
    \end{tikzpicture}}
\caption{\label{fig:sfun} Temperature structure functions of second (a,d,g), fourth (b,e,h) and tenth order (c,f,i)  from 2D RB convection at three different locations off the wall: $z^+=30$ (a-c), $z^+=100$ (d-f) and $z^+=200$ (g-i) for various Rayleigh numbers. The legend in (a) applies to all panels. Red dashed lines represent fits in the range $5z\leq r\leq 0.5L$, symbols mark $r/z =1$ (circles) and the extent of the fitting range (crosses), respectively, for later reference. }
\end{figure}

We begin by plotting structure functions of second, fourth and tenth order from 2D RB convection as a function of $r/z$ in figure \ref{fig:sfun}. For all cases, we present results at three distances from the wall, namely $z^+ =30$ (figure \ref{fig:sfun}a-c), $z^+=100$ (figure \ref{fig:sfun}d-f) and $z^+=200$ (figure \ref{fig:sfun}g-i), where as usual the superscript  $+$ indicates normalization  by $\nu/U_{\tau}$. Clearly, there is no scaling according to a scalar equivalent of (\ref{eq:Sdir}) for either $Ra$ at the position closest to the wall (figure \ref{fig:sfun}a-c). However, in the other cases an approximately linear region appears for $r/z>1$ making it tempting to fit the slopes directly. A fitting range $5z\leq r \leq 0.5 L$ (indicated by crosses in the figure) captures this region quite well for all $Ra$ and the resulting fits are shown as red dashed lines. No fits are computed at the lowest $Ra$ where the fitting range becomes prohibitively small.

\begin{figure} 
\psfrag{a}[c][c][1]{$(a)$}
\psfrag{b}[c][c][1]{$(b)$}
\psfrag{p}[c][c][1]{$p$}
\psfrag{Dp}[c][c][1]{$D^\theta_p$}
\psfrag{Ra}[c][c][1]{$Ra$}
\psfrag{DS1}[c][c][1]{$\Delta S^\theta_1/\Delta S^\theta_1(Ra = 10^{14})$}
\psfrag{zp1}[l][l][1]{$z^+=100$}
\psfrag{zp2}[l][l][1]{$z^+=200$}
\psfrag{ds}[l][l][0.9]{$D_p^u$, high-Re TBL}
\centering
\vspace{0.5cm}
\subfloat{\includegraphics[scale  =1]{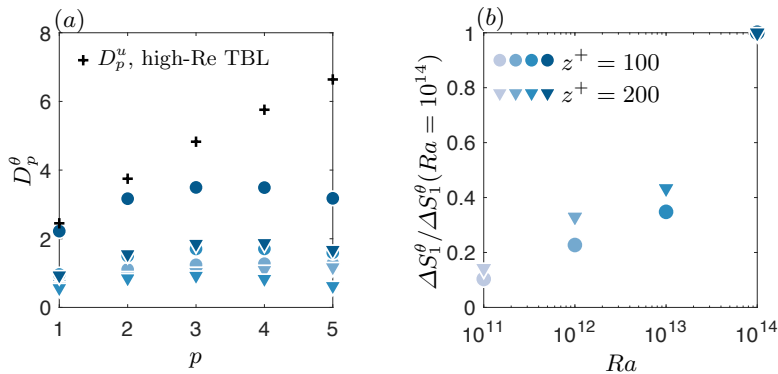}}
\caption{\label{fig:slopes}a) Slopes $D_p^\theta$ obtained from directly fitting the structure functions in the range $5z\leq r\leq 0.5L$ as indicated in figure \ref{fig:sfun} compared to high-$Re_{\tau}$ reference data from \citet{deSilva2015scaling}. b) $\Delta S_1^\theta$ as a function of $Ra$ normalized by  $\Delta S_1^\theta(Ra =10^{14})$. Colours indicate increasing $Ra$ from light to dark (see legend in figure \ref{fig:sfun}).}
\end{figure}

A detailed comparison of the values of $D_p^\theta$ obtained in this way  (figure \ref{fig:slopes}a) reveals that the results depend on both $Ra$ and $z^+$ in the investigated range. Generally, values of $D_p^\theta$ are higher at $z^+=100$ and this difference becomes larger with increasing $Ra$. In most cases, $D_p^\theta$ is very low and it is only at $z^+=100$ and  $Ra=10^{14}$ (and small $p$) that the values of $D_p^\theta$ are of comparable magnitude to results for $D_p^u$ in high-Re turbulent boundary layers (TBLs). We  point out that the decrease of $D_p^\theta$ with increasing $p$ observed in some cases at higher $p$ is unphysical and likely related to the insufficient fitting range at higher orders. We further emphasize that even at high $Re$ a direct match between $D_p^\theta$ and $D_p^u$ is not necessarily expected.  It is well established   \citep[see e.g.][for passive scalars]{Warhaft2000} that intermittency in the inertial range $\eta \ll r \ll z$  is higher for scalars as compared to the velocity itself. This translates to lower scaling exponents $\xi_{2p}$ in the corresponding scaling relation $S_1^\theta \sim (r/z)^{\xi^\theta_{2p}/p} $ as compared to the velocity counterpart $\xi_{2p}^u$. From matching the inertial scaling with (\ref{eq:Sdir}) at $r=z$ \citet{deSilva2015scaling} semi-empirically derived a linear relationship between $D_p^u$ and $\xi_{2p}^u$, suggesting that (disregarding other dependencies) these differences may persist in the energy-containing range investigated here.  However, a definitive answer to this question will have to be based on high-$Re$ data of the scalar field in wall-bounded turbulent flows.


\textcolor{black}{While the direct analysis of the slopes $D_p^\theta$ remains inconclusive, additional insight can be gained from studying the second order structure function $S_1^\theta$. In this case, the difference $\Delta S_1^\theta =S_1^\theta(r_2)-S_1^\theta(r_1) $ between the structure function at two different separation distances  $r_1$ and $r_2$ can be interpreted as the contribution of eddies with sizes in the range between $r_1$ and $r_2$ to the overall (scalar) energy. 
Consequently, when using the bounds of the log-linear scaling regime for $r_1$, $r_2$ as indicated in figure \ref{fig:sfun}d, this increment $\Delta S^\theta_1$, characterizes the  contribution of the energy-containing range to the overall energy. Note that here we adopted the bounds of the fitting region used above for  for simplicity, but other reasonable choices give qualitatively similar results.} Figure \ref{fig:slopes}(b), where $\Delta S^\theta_1$ is normalized by the result obtained at $Ra = 10^{14}$, shows that this quantity only increases mildly at low $Ra$. However, it rises steeply between $Ra =10^{13}$ and $Ra = 10^{14}$ consistent with transitional behaviour in this range. No significant differences arise between  $z^+=100$ and $z^+=200$ in this case.

\subsection{ Scalar ESS scaling in RB convection}\label{sec:ESS}

\begin{figure} 
\psfrag{a}[c][c][1]{$(a)$}
\psfrag{b}[c][c][1]{$(b)$}
\psfrag{c}[c][c][1]{$(c)$}
\psfrag{d}[c][c][1]{$(d)$}
\psfrag{e}[c][c][1]{$(e)$}
\psfrag{f}[c][c][1]{$(f)$}
\psfrag{g}[c][c][1]{$(g)$}
\psfrag{h}[c][c][1]{$(h)$}
\psfrag{i}[c][c][1]{$(i)$}
\psfrag{S1}[c][c][1]{$S_1^\theta$}
\psfrag{S2}[c][c][1]{$S_2^\theta$}
\psfrag{S3}[c][c][1]{$S_3^\theta$}
\psfrag{S5}[c][c][1]{$S_5^\theta$}
\centering
\vspace{0.5cm}
\subfloat{\begin{tikzpicture}
        \node[anchor=south west,inner sep=0] (image) at (0,0) {\hspace{0.45cm}\includegraphics[scale = 1]{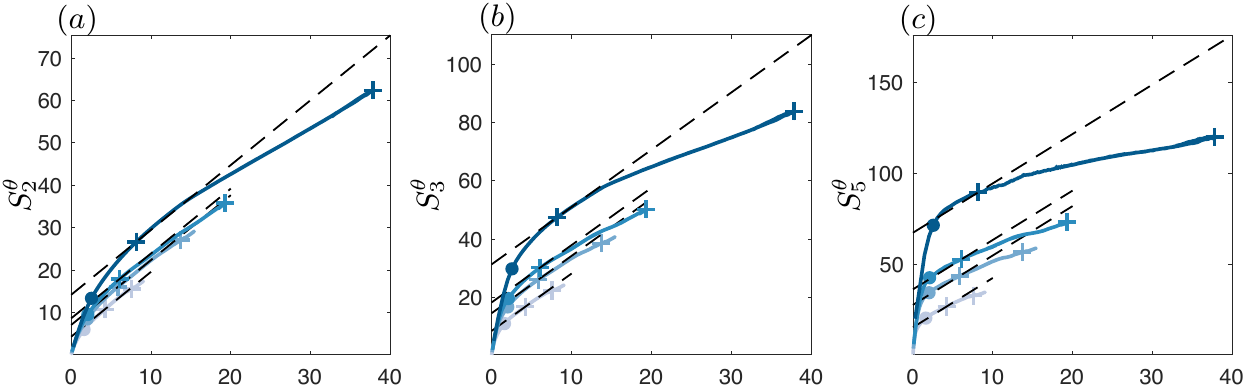}};
        \begin{scope}[x={(image.south east)},y={(image.north west)}]
                \node[inner sep=0,anchor=center] (note1) at (0,1.01) {\scalebox{0.9}{$z^+=30$}};
        \end{scope}
    \end{tikzpicture}}\\
    \subfloat{\begin{tikzpicture}
        \node[anchor=south west,inner sep=0] (image) at (0,0) {\hspace{0.4cm}\includegraphics[scale = 1]{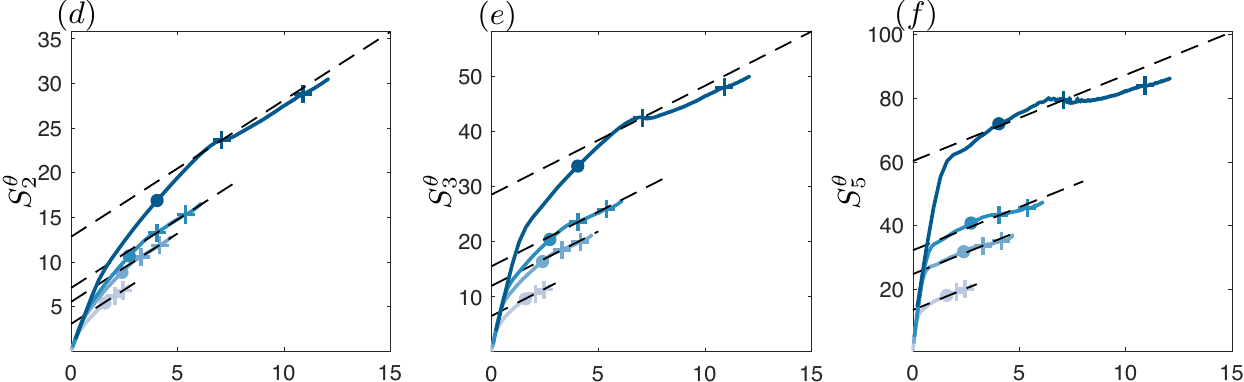}};
        \begin{scope}[x={(image.south east)},y={(image.north west)}]
                \node[inner sep=0,anchor=center] (note1) at (0,1.01) {\scalebox{0.9}{$z^+=100$}};
        \end{scope}
    \end{tikzpicture}}\\
    \subfloat{\begin{tikzpicture}
        \node[anchor=south west,inner sep=0] (image) at (0,0) {\hspace{0.35cm}\includegraphics[scale =1]{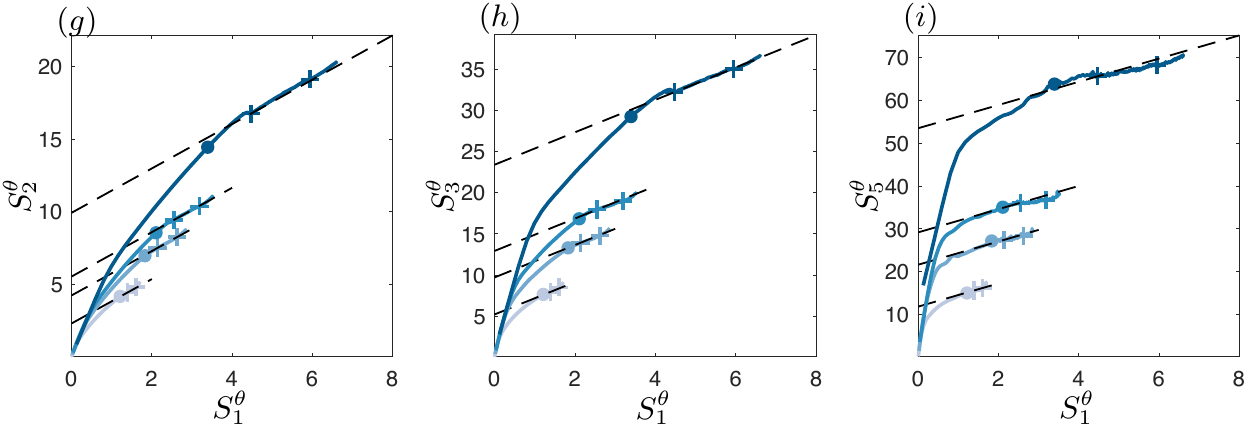}};
        \begin{scope}[x={(image.south east)},y={(image.north west)}]
                \node[inner sep=0,anchor=center] (note1) at (0,1.01) {\scalebox{0.9}{$z^+=200$}};
        \end{scope}
    \end{tikzpicture}}
\caption{ \label{fig:ESS}Scalar structure functions of from 2D RB convection higher order plotted versus $S_1$ at $z^+ = 30$ (a-c),$z^+ = 100$ (d-f), and $z^+ = 200$ (g-i) for various $Ra$ (legend of figure \ref{fig:sfun} applies). Dashed lines correspond to the relative slopes $D_p^u/D_1^u$ from \citet{deSilva2017} with the $y$-axis cutoff fit at $S_1(r=5z)$, symbols mark the locations of $r/z =1$ (circles) and the fitting region indicated in figure \ref{fig:sfun} (crosses) for reference.  }
\end{figure}

We now focus on the relative scalings $D_p^\theta/D_1^\theta$ according to (\ref{eq:ESSscalar}). The ESS framework has been demonstrated to extend the scaling regime not only to low $Re$ but also to a wider range of wall-normal distances. In particular, \citet{Krug2017ESS} found convincing scaling for $D_p^u/D_1^u$ as low as $z^+ =30$. From figure \ref{fig:ESS}a-c, it is clear that the same does not hold for the scalar structure functions in RB convection. Even at the highest $Ra$, there is no linear relationship between $S_1^\theta$ and $S_p^\theta$ for all orders considered at this location.
The situation improves at $z^+ = 100$ (figure \ref{fig:ESS}d-f), where particularly at the highest $Ra$ and low orders the curve begin to exhibit an approximately linear range. However, deviations become more apparent with increasing $p$ and results at the higher orders in figures \ref{fig:ESS}(e,f) demonstrate that the ESS scaling is not  yet fully attained at this position.
It is only at $z^+ = 200$ (figures \ref{fig:ESS}g-i) that a convincing ESS scaling according to (\ref{eq:ESSscalar}) is obtained up to tenth order. The scaling range is well established at $Ra = 10^{14}$, already significantly decreases in size at $Ra= 10^{13}$ and is basically non-existent at $Ra=10^{11}$. This behaviour is in very good correspondence to the changes in the $Nu$ scaling in figure \ref{fig:Nu}a,  corroborating that the change in scaling observed there is indeed due to a transition in the  BL structure from laminar (no scaling) to turbulent (with ESS scaling). It should be noted that, just as for velocity structure functions,  ESS scaling at $z^+ = 200$ is observed for $r/z \gtrapprox 1$ for all $Ra$, i.e. the spatial scaling range is the same. However, consistent with the results in figures \ref{fig:sfun} and \ref{fig:slopes}(b), there is hardly any energy in this range at $Ra$ below transition to the ultimate regime. Remarkably, the relative slopes  attained for the temperature structure functions in 2D RB at $z^+ = 200$  appear to be the same as those measured for their velocity counterparts and reported in \citet{deSilva2017}. This is remarkable since as pointed out in \S\ref{sec:sfun}, the directly measured slopes $D_p^\theta$ and $D_p^u$ need not be the same. So this feature seems to be one of scalar fields.

\subsection{Scalar ESS scaling in channel flow}\label{sec:ESSch}

\begin{figure} 
\psfrag{a}[c][c][1]{$(a)$}
\psfrag{b}[c][c][1]{$(b)$}
\psfrag{c}[c][c][1]{$(c)$}
\psfrag{d}[c][c][1]{$(d)$}
\psfrag{e}[c][c][1]{$(e)$}
\psfrag{f}[c][c][1]{$(f)$}
\psfrag{S1}[c][c][1]{$S_1^\theta$}
\psfrag{S2}[c][c][1]{$S_2^\theta$}
\psfrag{S3}[c][c][1]{$S_3^\theta$}
\psfrag{S5}[c][c][1]{$S_5^\theta$}
\psfrag{ronz}[c][c][1]{$r/z$}
\psfrag{zp30}[l][l][0.8]{$z^+ =30$}
\psfrag{zp100}[l][l][0.8]{$z^+ =100$}
\psfrag{zp200}[l][l][0.8]{$z^+ =200$}
\centering
\vspace{0.5cm}
\subfloat{\includegraphics[scale = 1]{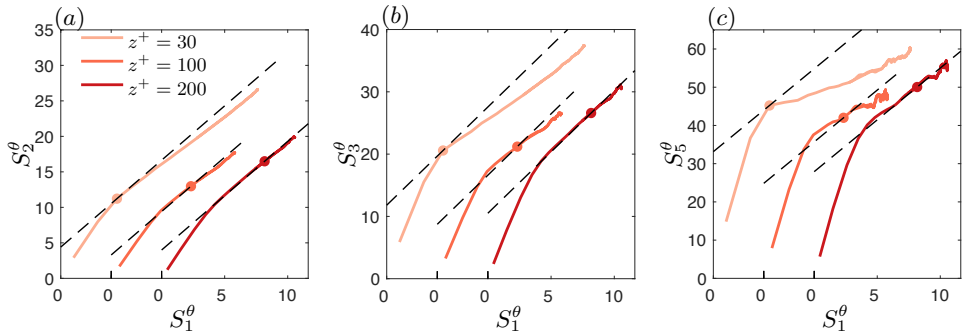}}
\caption{\label{fig:chan} Structure function results for a passive scalar in turbulent channel flow at $Re_{\tau}=590$ in ESS form at fourth (a), sixth (b) and tenth order (c) relative to $S_1^\theta$. Results at different $z^+$ are shifted by 4 on the $x$-axis for clarity. Dashed lines correspond to the relative slopes $D_p^u/D_1^u$ from \citet{deSilva2017}  and circles indicate the location where $r=z$ for reference. }
\end{figure}

The question remains why ESS scaling is only observed at a larger distance from the wall in the present case as compared to previous findings for velocity structure functions. At this point, possible explanations to be considered are that this might be either a consequence of the 2D setup, a property of RB convection or a feature of the scalar field. To address this, we present ESS results of scalar structure functions in turbulent channel flow in figure \ref{fig:chan}(a-c) at different orders. Indeed, the results are very similar to what was observed for 2D RB before in that there is no ESS scaling for $z^+=30$ and $z^+=100$. And again ESS scaling is recovered at $z^+=200$ with relative slopes matching those measured for velocity structure functions mirroring the observations made for RB convection. 

\section{Discussion and Conclusions}\label{sec:conc}
We have analysed the temperature boundary layers in the energy-containing range of 2D RB convection by means of temperature structure functions. Even though the RB structure functions exhibit log-linear scaling for $r/z\gtrapprox 1$ when plotted against separation distance, the slopes remain small at low $Ra$ and they were found to vary significantly with both $Ra$ and $z^+$, rendering the analysis inconclusive in this point. While a dependence of the slopes on $Re_{\tau}$ and wall-normal position is also observed for velocity structure functions \citep[see][]{Krug2017ESS}, typically the log-linear scaling is much less evident in these cases compared to figure \ref{fig:sfun} at $z^+\geq 100$. 
Also for the scalar in channel flow investigated here (plots not shown) a direct scaling regime is not discernible such that it appears likely that the more prominent log-linear regimes in figure \ref{fig:sfun} are a consequence of the 2D setup. An important point remains however, that the contribution of the energy-containing range to the total energy increases significantly beyond $Ra = 10^{13}$, which coincides with the transition in $Nu$ vs. $Ra$ scaling.

The main finding of the paper is the clear evidence that the temperature structure function in the BLs of turbulent RB flows exhibit ESS scaling in the energy-containing range for large enough wall distances $z^+\gtrapprox 100$. The extent of the scaling range thereby reflects the behaviour of $Nu$ very  well in increasing from non-existent at $Ra = 10^{11}$ (well in the classical regime) to considerable beyond the $Nu$ scaling transition at $Ra = 10^{14}$. This provides further evidence that said transition is indeed related to a switch from laminar type to turbulent type thermal BLs.

Moreover, we establish that the relative slopes for scalar structure functions in the ESS-form are the same as those previously obtained when analysing velocity BLs. This is confirmed by comparing the RB results to those obtained in a planar channel geometry. In both cases, ESS scaling is only established at $z^+= 200$ which is different from velocity structure functions and appears to be a feature of the scalar field. \textcolor{black}{Interestingly enough, already \citet{Perry1982}  point out differences between the scalar and the velocity field. They argue that at the end of the `lifetime' of an eddy, the vorticity contributions in the two rods of the assumed hairpin cancel such that no induced velocity field remains. However, such a cancellation does not apply for the scalar transported by the eddy, such that  the `debris', as they call it, of past eddies  sets up a scalar background profile.}  If and how exactly this is related to the observations made here remains however unclear.

At least for the present cases, there also appears to be no difference between active (in RB flow) and passive (in the channel flow) scalars.  In other regards, the deviations close to the wall underline in addition to the fact the observed values of $D_p/D_m$ are sub-Gaussian \citep{Krug2017ESS} that the ESS scaling and the values of the relative slopes are indeed non-trivial.

\acknowledgements{
The work was financially supported by
NWO-I and  the Netherlands Center for
Multiscale Catalytic Energy Conversion (MCEC), both
sponsored by the Netherlands Organization for Scientific
Research (NWO). Part of
the simulations were carried out on the Dutch national
e-infrastructure with the support of SURF Cooperative.
We also acknowledge PRACE for awarding us access to
Marconi based in Italy at CINECA under PRACE Project
No. 2016143351 and the DECI resource Archer based in
the United Kingdom at Edinburgh with support from the
PRACE aisbl under Project No. 13DECI0246. We also acknowledge the support of the Australian Research Council and the McKenzie Fellowship program at the University of Melbourne.}

\bibliographystyle{jfm}
\bibliography{jfm_ess_rb}

\end{document}